\documentclass{article}
\usepackage[a4paper]{geometry}
\usepackage{amssymb,amsmath,amsfonts,amsthm,tikz,braket,hyperref}

\newcommand{\openone}{\leavevmode\hbox{\small1\kern-3.8pt\normalsize1}}
\newcommand{\ketbra}[2]{\ket{#1}\hspace*{-.75mm}\bra{#2}}
\newcommand{\av}[1]{\ensuremath{\langle #1 \rangle}}
\newcommand{\M}{\ensuremath{\mathbb{M}}}
\newcommand{\A}[1][n]{\ensuremath{\M_{#1}}}
\renewcommand{\H}{\ensuremath{\mathbb{H}}}

\newcommand{\R}{\ensuremath{\mathbb{R}}}

\newcommand{\be}{\begin{equation}}
\newcommand{\ee}{\end{equation}}
\newcommand{\bea}{\begin{eqnarray}}
\newcommand{\eea}{\end{eqnarray}}
\newcommand{\chan}[2]{\mathcal{E}_{#2 | #1}}
\newcommand{\chancomma}[2]{\mathcal{E}_{#2|#1}}
\newcommand{\cs}[2]{E_{#2 | #1}}
\newcommand{\cscomma}[2]{E_{#2|#1}}
\DeclareMathOperator{\tr}{Tr}

\theoremstyle{plain}
\newtheorem{theorem}{Theorem}
\newtheorem{lemma}[theorem]{Lemma}

\begin{document}
\title{Can a quantum state over time resemble a quantum state at a single time?}
\author{Dominic Horsman\thanks{Department of Physics, Durham, University, United Kingdom}
\and Chris Heunen\thanks{School of Informatics, University of Edinburgh, Scotland}
\and Matthew F. Pusey\thanks{Perimeter Institute for Theoretical Physics, Waterloo, Canada};
\and Jonathan Barrett\thanks{Department of Computer Science, University of Oxford, United Kingdom}
\and Robert W. Spekkens\footnotemark[3]}
\maketitle

\begin{abstract}
  The standard formalism of quantum theory treats space and time in fundamentally different ways. In particular, a composite system at a given time is represented by a joint state, but the formalism does not prescribe a joint state for a composite of systems at different times. If there were a way of defining such a joint state, this would potentially permit a more even-handed treatment of space and time, and would strengthen the existing analogy between quantum states and classical probability distributions. Under the assumption that the joint state over time is an operator on the tensor product of single-time Hilbert spaces, we analyze various proposals for such a joint state, including one due to Leifer and Spekkens, one due to Fitzsimons, Jones and Vedral, and another based on discrete Wigner functions. Finding various problems with each, we identify five criteria for a quantum joint state over time to satisfy if it is to play a role similar to the standard joint state for a composite system: that it is a Hermitian operator on the tensor product of the single-time Hilbert spaces; that it represents probabilistic mixing appropriately; that it has the appropriate classical limit; that it has the appropriate single-time marginals; that composing over multiple time-steps is associative. We show that no construction satisfies all these requirements. 
  If Hermiticity is dropped, then there is an essentially unique construction that satisfies the remaining four criteria.
\end{abstract}

\section{Introduction}\label{sec:intro}

Quantum theory, as usually formalised, contains a fundamental asymmetry between space and time. This is evident when considering the description of composite quantum systems. If a composite system consists of several components existing at a given time, then the formalism specifies that the joint state is given by a density matrix acting on the tensor product of the Hilbert spaces associated with the components. (This is the case regardless of whether the components are spatially separated systems or different degrees of freedom of a single system.) But in principle, there is another sort of composite system besides these, namely one where the components are time-like separated. In this case, there is no standard prescription for the joint state of the composite. Rather, states are defined only at a single time, and evolve over time under the action of a Hamiltonian. 
But is this asymmetry fundamental or merely an artefact of a particular formalism?  Relativity theory has revealed that many distinctions between space and time previously thought to be fundamental are not.  While spatial and temporal dimensions do have opposite signs in the metric and causal ordering is preserved under Lorentz transformations, the distinction between spatial and temporal intervals becomes a merely observer-dependent decomposition of the  fundamental concept of a spatio-temporal interval.  This naturally prompts the question of whether the distinction in quantum theory between how one models composite systems at a single time and how one models composite systems over time is like the distinction between spatial and temporal intervals, and can be eliminated, or whether it is like the sign distinction in the metric and cannot. 
The asymmetry is also in marked contrast with classical probability theory, where joint probabilities can be defined for sets of events whatever their spatio-temporal relationships. Hence although there is a well developed analogy between density matrices and classical probability distributions, the analogy is limited in scope when it comes to time-like separated systems.

If the apparent asymmetry between space and time in quantum theory is fundamental, then the understanding of time given by quantum theory must be different from that suggested by a combination of relativity and classical probability theory. If, on the other hand, it is not fundamental then it should be removable. One way of removing the asymmetry would be to construct quantum states for composites over time. States would then be defined across both space and time, without a separate formalism describing evolution over time. 

There have been a variety of proposals for expressing quantum theory in a manner that treats space and time in a more even-handed fashion. The sum-over-histories approach to quantum dynamics~\cite{feynman,Sorkin1997} and the program of consistent/decoherent histories~\cite{griffiths1984consistent,omnes1988logical,GellmannHartle1990} are examples.
An alternative family of proposals includes the multi-time formalism (see~\cite{aharonov2007two, aharonov2009multiple} and follow-up work~\cite{aharonov2013each, silva2014pre}), quantum combs (see~\cite{ChiribellaEtAl_2009}, follow-up work~\cite{Chiribella_2012, ChiribellaEtAl_2013} and related formalisms \cite{gutoskiwatrous:games,oreshkovcerf:withouttime}), process matrices (see~\cite{Oreshkov2012} and follow-up work~\cite{Costa2016}) and the causaloid formalism (see~\cite{hardy2007towards,Hardy3385}). 
These latter proposals, although differing from one another in notation and in the types of problems that have been addressed, share the feature that a quantum system at a localised region of space-time is associated with two Hilbert spaces, one carrying an incoming state, and one an outgoing state. Part of the reason for this is that each proposal allows for the possibility that an agent situated at that region of space-time can intervene upon the system, with the intervention corresponding to a quantum instrument: the quantum instrument is a set of trace non-increasing completely positive maps, one for each classical outcome of the intervention, that mediate the incoming and outgoing Hilbert spaces.

In such approaches, the main criterion of success is whether the formalism can be used to compute 
the joint probability distribution over the outcome variables for a given set of interventions.
Our project, by contrast, is also motivated 
by the goal of providing a {\em causal account} of the operational predictions.  To do so,  it is critical to have a formulation of quantum theory that makes a clean distinction between the aspects of the formalism that are about causal influence and those that are about Bayesian inference.
For instance, rather than merely predicting a correlation between the outcomes of two measurements, such a formulation specifies whether this correlation is due to a common cause of the two variables or rather a cause-effect relation between them~\cite{Ried2015}.   One proposal for how to achieve this separation takes joint quantum states to be inferential objects, the quantum analogues of joint probability distributions~\cite{leiferspekkens:bayesian}.  Given that a joint probability distribution is the appropriate way to describe an agent's incomplete information of a composite system, {\em regardless} of the spatio-temporal relations that hold among its components, it is natural to ask whether there is a notion of a joint quantum state that is similarly applicable to an arbitrary composite and which coincides with the standard notion when the composite consists of a set of systems considered at a given time.  We focus on a special case of this question: whether there is a notion of a joint quantum state for a composite over time which mirrors that of a joint quantum state for a composite at a single time.

If such a project could succeed, not only would it extend and strengthen the analogy between quantum states and classical probability distributions, it would also lend support to the view that a quantum state can be thought of as information directly about a system as opposed to a form of information that can only be expressed in terms of outcomes of potential measurements on the system. 
That is, it would lend support to the view that the quantum state represents a state of knowledge of an underlying reality associated to the system as opposed to merely representing a state of knowledge of the outcomes of measurements that one might implement on the system.
It would also shed light on the subject of the quantum-classical correspondence: if a subset of the systems in a network describe measurement pointers (and are therefore suitably macroscopic and decohered) then the joint quantum state for this subset would encode the joint probability distribution over the pointer variables.  Such a formulation could be of practical use for the analysis of quantum information-processing protocols.
It could also help to pinpoint where any remaining asymmetry between space and time arises in the quantum formalism.

We therefore undertake to investigate the possibility of defining quantum joint states over time in a manner that closely matches the standard quantum treatment of composite systems at a given time.  In particular, we assume that a temporally localised $d$-dimensional quantum system is associated with a single $d$-dimensional Hilbert space (rather than a pair of such Hilbert spaces, as is assumed in Refs.~\cite{aharonov2007two, aharonov2009multiple, aharonov2013each, silva2014pre,ChiribellaEtAl_2009, Chiribella_2012, ChiribellaEtAl_2013,gutoskiwatrous:games,oreshkovcerf:withouttime,Oreshkov2012, AraujoEtAl_2014, Costa2016,hardy2007towards,Hardy3385}), and that the state of a composite system over time is an operator on the tensor product of the Hilbert spaces associated with the temporally localised components.
  We explore a number of definitions along these lines, including a proposal of Leifer and Spekkens (LS)~\cite{leiferspekkens:bayesian} (inspired by the view that quantum theory is a generalisation of classical probability theory), a proposal of Fitzsimons, Jones and Vedral (FJV)~\cite{fitzsimonsjonesvedral:causation}, and a novel proposal based on discrete Wigner functions. Following a close examination of these different options, we distill five criteria that the state of a composite over time should satisfy if it is to have the essential qualities of the state of a composite at a single time. Our main theorem states that there is no way to construct a state over time that satisfies all of these criteria. We also show that if one of the criteria is dropped -- namely, the criterion that the state over time is a Hermitian operator -- then the only construction satisfying the remaining four criteria corresponds to straightforward matrix multiplication of the initial state and the operators representing the quantum channels that evolve from one time-slice to the next. Finally, we show that if Hermiticity is retained, but an assumption of associativity is dropped, then the FJV construction satisfies the remaining criteria.  

\section{States over time}\label{sec:states}

\begin{figure}[t]
    \scalebox{0.75}{\hspace{2cm}
     \begin{minipage}[c]{1.0\linewidth}
     \centering
     \[\begin{tikzpicture}
       \node[circle,draw] (a) at (0,0) {A};
       \node[circle,draw] (b) at (0,2) {B};
       \draw[line width=1.5pt,->] (a) -- (b);
       \draw (-1,0) node {$\rho_A$};
       \draw (-1,1) node {$\chan{A}{B}$};
       \draw (-1,2) node {$\rho_B$};
     \end{tikzpicture}\]
	\end{minipage}}
  \caption{Two quantum systems $A$ and $B$, and a channel between them. A potential quantum state over time $\rho_{AB}$ should be a function of these variables.} \label{first}
\end{figure}
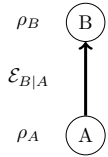

We begin with the simplest situation in which to discuss the construction of a quantum state over time: a single system considered at two different times, as depicted in Fig.~\ref{first}. 
Whereas Hilbert spaces are conventionally associated with systems that persist over time, we shall here associate Hilbert spaces to quantum systems that are considered to be localized to a given region of space-time (similar in spirit to `events' in relativity). The system at the first time is denoted $A$, and that at the second time $B$. Assuming that the system is initially uncorrelated with its environment, the most general possible evolution of the system corresponds to a \emph{quantum channel}, that is, a trace-preserving completely positive map, denoted $\chan{A}{B}$. Conventionally, the quantum state of $A$ is represented by an operator acting on $\mathcal{H}_A$, denoted $\rho_A$, and the quantum state of $B$ is represented by an operator acting on $\mathcal{H}_B$, denoted $\rho_B$. The state of $B$ can be determined by the state of $A$ and the channel, via $\rho_B = \chan{A}{B}(\rho_A)$. 

It is natural to consider the possibility of a quantum state for the composite system $AB$, represented by an operator $\rho_{AB}$ acting on $\mathcal{H}_{A}\otimes \mathcal{H}_{B}$. The scenario of Fig.~\ref{first} is fully defined by the state $\rho_A$ and the channel $\chan{A}{B}$ (as these determine the output state $\rho_B$), so we assume that the state over time is a function of these:
\be \rho_{AB} = f(\rho_A,\chan{A}{B}) \label{func}.\ee
We restrict to the finite-dimensional case, so that $\rho_{AB}$ can be represented by a square matrix, but we do not require that $\rho_{AB}$ is a standard density matrix. The properties that should be required of the function $f$, and the possible forms $f$ can take, are the main questions addressed in this paper.

\subsection{Three possible definitions of a state over time}

It will be useful in the following to represent the channel $\chan{A}{B}$ by an operator acting on $\mathcal{H}_A \otimes \mathcal{H}_B$, denoted $\cs{A}{B}$.  As it is also useful if the representation of the channel  is basis-independent,   we choose it to be the operator that is Jamio{\l}kowski-isomorphic~\cite{Jamiolkowski1972} to $\chan{A}{B}$, defined as
\[
\cs{A}{B} \equiv \sum_{ij} \chan{A}{B} ( \ketbra{i}{j}_A ) \otimes \ketbra{j}{i}_A,
\]
where $\{ |i\rangle \}$ is any orthonormal basis of $\mathcal{H}_A$.   
(Up to normalisation, this operator is the partial transpose of the operator that is Choi-isomorphic~\cite{choi} to $\chan{A}{B}$.) The operator $\cs{A}{B}$ is independent of the basis used in its definition,  and the state of the output of the channel can be expressed in terms of it as~\cite{leiferspekkens:bayesian}: 
\[
\rho_B = \mathrm{Tr}_A( \cs{A}{B} \rho_A ).
\] 
We call $\cs{A}{B}$ the \emph{channel state}. 

The joint state $\rho_{AB}$ can be understood as being given by a binary operation $\star$ that takes a pair of operators on $\mathcal{H}_A\otimes \mathcal{H}_B$ to an operator on $\mathcal{H}_A \otimes \mathcal{H}_B$:
\be\label{star0}
\star \colon (\cs{A}{B}, \rho_A \otimes \openone_B)\mapsto \rho_{AB}.
\ee
We call this operation a \textit{star product}. For notational convenience, we will in the following often suppress tensor products with identity, writing for example $\rho_A$ instead of $\rho_A\otimes \openone_B$, so that
\be \rho_{AB} = \cs{A}{B} \star \rho_A. \label{star}\ee 

There have been a number of previous attempts to define a joint state $\rho_{AB}$ along these lines. The proposal due to Leifer and Spekkens (LS) constructs joint states over time in the context of a programme of reframing quantum theory as a theory of Bayesian inference \cite{leiferspekkens:bayesian},   which builds on \cite{leifer2006quantum} and \cite{leifer} . Just as a density matrix on a composite of $A$ and $B$, $\rho_{AB}$, can be understood to be a quantum analogue of a joint probability distribution, $P(XY)$,   LS observe that a channel state $\cs{A}{B}$ can be understood as a quantum analogue of a conditional probability distribution, $P(Y|X)$, which describes a classical channel with input $X$ and output $Y$ .
For example,  the analogue of the normalization condition for conditional probabilities, $\sum_Y P(Y|X) = 1$, is $\mathrm{Tr}_B( \cs{A}{B} ) = \openone_A$, and the analogue of the fact that the output of a classical channel satisfies $P(Y) = \sum_X P(Y|X) P(X)$ is the fact that the channel state satisfies $\rho_B = \mathrm{Tr}_A( \cs{A}{B} \rho_A )$. In the classical case, there is no obstacle to defining a joint probability distribution for the pair of systems constituting the input and the output of a channel via $P(XY) = P(Y|X)P(X)$. In the quantum case, however, the operators $\cs{A}{B}$ and $\rho_A\otimes \openone_B$ do not in general commute, so it is not obvious what should replace this equation.  LS considered the following option:
\be \rho^{(\mathrm{LS})}_{AB} = \cs{A}{B} \star_{\mathrm{LS}} \rho_A 
\equiv \rho_A^\frac{1}{2} \  \cs{A}{B} \  \rho_A^\frac{1}{2}\label{ls},\ee
where the right-hand side is a shorthand notation for $ (\rho_A^\frac{1}{2}\otimes \openone_B) \  \cs{A}{B} \  (\rho_A^\frac{1}{2}\otimes \openone_B)$. The operator $\rho^{(\mathrm{LS})}_{AB}$ is Hermitian, and although it is not in general positive, it is locally positive,
meaning that $\bra{ab} \rho^{(\mathrm{LS})}_{AB} \ket{ab} \geq 0$ for all $\ket{a}$ in $\mathcal{H}_A$ and $\ket{b}$ in $\mathcal{H}_B$.    Significant problems with this proposal were already noted by LS in Sec.~VII of \cite{leiferspekkens:bayesian}. 

An alternative proposal for a state over time is that of Fitzsimons, Jones, and Vedral (FJV) \cite{fitzsimonsjonesvedral:causation}. As it stands, the proposal is for multi-qubit systems, rather than systems of arbitrary dimension. In this work, it will suffice to consider the FJV construction for the special case in which $A$ and $B$ are qubits. In order to define the FJV joint state,  imagine that a Pauli measurement $\sigma_i$ is performed on $A$, with outcome $\pm 1$ (here, the Pauli operators are denoted $\sigma_0 = \openone, \ \sigma_1=\sigma_x, \ \sigma_2=\sigma_y, \ \sigma_3=\sigma_z$). After this measurement, the state of $A$ is imagined to be updated according to the projection postulate, before evolving according to the channel $\chan{A}{B}$, and a Pauli measurement $\sigma_j$ is then performed on $B$. Let $\av{\sigma_i \otimes \sigma_j} $ denote the expectation value of the product of the outcomes of the two measurements. The FJV joint state is given by
\be \rho^{(\mathrm{FJV})}_{AB} \equiv \frac{1}{4} \left( \sum_{i,j=0}^{3}\av{\sigma_i \otimes \sigma_j} \ \sigma_i \otimes \sigma_j  \right). \label{fjv}\ee
Note that although the FJV joint state is defined by reference to measurements that one imagines to be performed on $A$ and on $B$, the joint state is intended as a description of the system $AB$ in the 
case where no measurements are made on either, but rather $A$ simply evolves into $B$ according to the given channel. The matrix $\rho^{(\mathrm{FJV})}_{AB}$ is Hermitian, but not necessarily positive   (nor even necessarily locally positive).  It is clear that $\rho^{(\mathrm{FJV})}_{AB}$ depends only on $\cs{A}{B}$ and $\rho_A$, hence can in principle be written in the form of Eq.~(\ref{star}), for a suitable definition of the star product. In fact, as we show in Appendix~\ref{sec:fjv}, the FJV construction for two qubits corresponds to the Jordan product:
\be 
\rho^{(\mathrm{FJV})}_{AB} =  \cs{A}{B} \star_{\mathrm{FJV}} \rho_A =\frac12\left( \rho_A \cs{A}{B} + \cs{A}{B}\rho_A \right). \label{fjvstar}
\ee
For  a single qubit considered 
at an arbitrary number of time steps, the FJV construction is similar, involving imagined Pauli measurements performed 
at each time step, and coincides with an iterated application of the Jordan product. 

Finally, one might also hope to construct a state over time in a manner that is analogous to the definition of a joint probability distribution,   but where probabilities are replaced by the {\em quasi-probabilities} appearing in a discrete Wigner representation of the states and channels \cite{wooters:discretewigner, Gross}. 
  A discrete Wigner (W) representation for a $d$-dimensional system $A$ is defined by a set $\Omega^A = \{ K^A_i\}$ of \emph{phase-point operators} which form a basis for the space of operators on $\mathcal{H}_A$, satisfying $\mathrm{Tr} (K^A_i K^A_j) = d \delta_{ij}$ and $\sum_i K^A_i = d \openone_A$, hence $\mathrm{Tr}(K^A_i) = 1$. A density matrix $\rho_A$ can be written $\rho_A = \sum_i r^A(i) K^A_i$, where $r^A(i)$ is a real-valued function on $\Omega^A$, with $\sum_i r^A(i) = 1$. Hence a system $A$ can be described by the function $r^A(i) \in [-1,1]$, which has the form of a quasi-probability distribution on a discrete phase space. Similarly, the operator $\cs{A}{B}$ representing a channel can be written $\cs{A}{B} = \frac{1}{d} \sum_{ij} r^{B|A}(j|i) K^A_i \otimes K^B_j$, where $r^{B|A}(j|i)$ is a real-valued function on $\Omega^A \times \Omega^B$. The extra factor of $1/d$ is introduced so that $r^{B|A}(j|i)$ satisfies $\sum_j r^{B|A}(j|i) = 1$; these can therefore be thought of as conditional quasi-probabilities. The natural definition for the W representation of the state of the composite $AB$ is then $r^{AB}(ij) \equiv r^{B|A}(j|i) r^A(i)$. This implies that the state over time is represented by the operator
\be \rho^{(\mathrm{W})}_{AB} \equiv  \sum_{ij} r^{B|A}(j|i) r^A(i) K^A_i \otimes K^B_j, \ee
which, when expressed explicitly as a function of $\rho_A$ and $\cs{A}{B}$, has the form 
\be
  \rho^{(\mathrm{W})}_{AB} 
  =  \cs{A}{B} \star_{\mathrm{W}} \rho_A 
  \equiv \frac{1}{d^2} \sum_{ij} \tr_{AB}(  \cs{A}{B} K^A_i \otimes K^B_j) \tr_A( \rho_A K^A_i) K^A_i \otimes K^B_j. \label{dw}
\ee

There are various ways, then, in which one might aim to define quantum states over time. This leads us to pose the question: can we isolate specific axioms or desiderata that a state over time ought to satisfy? Furthermore, given those axioms, is it possible to find a construction that satisfies them all? 

\section{Five criteria for a star product}\label{sec:criteria}

We introduce five basic axioms for a star product, motivating each by a corresponding desideratum for the properties of a state over time.

\subsection{Hermiticity}

We assume that, for a finite-dimensional system considered at $k$ distinct time steps, the state over time, $\rho_{A_1\cdots A_k}$, is a Hermitian operator   on the $k$-fold tensor product of copies of its Hilbert space, $\mathcal{H}_{A_1}\otimes\cdots\otimes\mathcal{H}_{A_k}$.  We denote the space of such operators by $\H_{A_1 \dots A_k}$, so that $\rho_{A_1\cdots A_k} \in \H_{A_1 \dots A_k}$.
In particular,   if $A$ and $B$ denote a system at two times, then $\rho_{AB} \in \H_{AB}$.  

Given that $\rho_A \otimes \openone_B \in \H_{AB}$ and $E_{B|A}\in \H_{AB}$, it follows that we are assuming that the star product of Eq.~\eqref{star0} is a map of the form: 
\[
\star \colon \H_{AB} \times \H_{AB} \to \H_{AB}.
\]
As mentioned, all three constructions, LS, FJV, and W, satisfy the assumption of Hermiticity.
Below, we consider the consequences of dropping it.

\subsection{Preservation of probabilistic mixtures}

The output state of a quantum channel for a given input state is a linear function of both the input state and the channel.  This is necessary to ensure that a probabilistic mixture of input states maps under the channel to the corresponding mixture of output states, and that a probabilistic mixture of channels acting on a given input state yields the corresponding mixture of output states.  By analogy, we would like the joint state over time for a probabilistic mixture of input states to be the corresponding mixture of the joint states for each input state, and similarly for probabilistic mixtures of channels.

Consider as an example the situation in Fig.~\ref{first}, but where the quantum system $A$ is conditioned on a classical variable $x$.    For instance, it could be that $x$  is the outcome of a fair coin toss, and a qubit $A$ is prepared in a state that depends on this outcome: $\rho_{A,x=h} \equiv \ketbra{0}{0}$ for heads, and $\rho_{A,x=t} \equiv \ketbra{1}{1}$ for tails.   If the channel between $A$ and $B$ is the identity channel, then $\cs{A}{B}^{\rm (id)} \equiv 2\ketbra{\phi^+}{\phi^+}^{T_B}$ where $\ket{\phi^+}\equiv \frac{1}{\sqrt{2}}\left( |0\rangle|0\rangle +|1\rangle|1\rangle \right)$ and where $T_B$ denotes the partial transpose on $B$. In this case, we would like the star product to satisfy 
\be
  \cs{A}{B}^{\rm (id)} \star \Big{(} \frac{1}{2}\rho_{A,x=h}  + \frac{1}{2}\rho_{A,x=t}  \Big{)} 
  =  \frac{1}{2}\left( \cs{A}{B}^{\rm (id)} \star \rho_{A,x=h} \right) + \frac{1}{2}\left( \cs{A}{B}^{\rm (id)} \star \rho_{A,x=t} \right). \label{notlinear}
\ee

In general, we assume:
\be
  \textbf{Convex-bilinearity}:
  \left\{  \begin{array}{c} 
  \big(px + (1-p)y\big) \star z = p(x \star z) + (1-p)(y \star z), \\
  x \star \big(py + (1-p)z) = p(x \star y) + (1-p)(x \star z).
  \end{array} \right. 
\ee
 	
While the FJV and W formulations are both straightforwardly convex-bilinear, there is a failure of this axiom in the case of the LS star product, which does not preserve convex combinations in $\rho_A$. For example, using the classical coin toss example, it is easy to show that the LS star product yields the mixture $\frac{1}{2} \ketbra{00}{00} + \frac{1}{2} \ketbra{11}{11}$ for the right-hand side of \eqref{notlinear}, but the non-separable $\ketbra{\phi^+}{\phi^+}^{T_B}$ for the left-hand side.

\subsection{Preservation of classical limit}
In classical probability theory, defining joint probabilities for events at different times is unproblematic. This gives a constraint on the form of quantum temporal joint states, if it is assumed that classical joint probabilities must be reproduced when a quantum system is behaving entirely classically. For the situation of Fig.~\ref{first}, this is the case if both the input quantum state and the operator corresponding to the channel are diagonal in the same basis. Consider for example, a channel that is completely dephasing in some basis,  and an input state that is diagonal in the same basis. For some probabilities $p(i)$ and conditional probabilities $p(j|i)$, the input state can be written $\rho_A = \sum_i p(i)\ketbra{i}{i}_A$, and the behaviour of the channel is given by 
\be \mathcal{E}_{B|A} (\rho_A) = \sum_i \braket{i|_A \rho_A |i}_A \sum_j p(j|i) \, \ketbra{j}{j}_B. \ee
For this channel, 
\be 
\cs{A}{B} = \sum_{i,j} p(j|i) \, \ketbra{i}{i}_A \otimes \ketbra{j}{j}_B , 
\ee
which is simply a matrix encoding of the classical conditional probabilities that define the channel. The classical joint probabilities would be given by $p(i,j) = p(i)p(j|i)$, which, when encoded in a matrix in the same way, are given by 
\be 
\rho_{AB} =\sum_{i,j} p(j|i) p(i) \, \ketbra{i}{i}_A \otimes \ketbra{j}{j}_B = \rho_A \cs{A}{B}.
\ee

In order to reproduce classical probability theory, therefore, when states and channel states are simultaneously diagonalisable the joint state over time should be given by their matrix product.  This is ensured if
\be [\cs{A}{B},\rho_A] = 0 \implies \cs{A}{B} \star \rho_A = \cs{A}{B} \rho_A .\ee
Hence we assume that a star product satisfies 
\be
  \textbf{Product on commuting pairs}: [x,y] = 0 \implies x\star y = xy.
\ee
Note that an implication of this condition is that composition with the identity satisfies
\be x\star \openone = \openone \star x = \openone x = x , \label{identity}\ee
since $[x,\openone]=0$ for all operators $x$.

It is easy to see that the LS joint state satisfies $\rho^{LS}_{AB} = \cs{A}{B}\rho_A$ when $\cs{A}{B}$ and $\rho_A$ commute.    This is also the case for the FJV joint state for a qubit at two times, as is evident from the form of Eq.~(\ref{fjvstar}). However, the W construction does not satisfy this axiom. Consider, for example, an input state that is a pure state of a qubit, and a completely dephasing channel that preserves this state. Direct calculation shows that unless the input state is an eigenstate of one of the Pauli operators, the Wigner joint state $\rho^{(W)}_{AB} \ne E_{B|A}\rho_A$.

\subsection{Preservation of marginal states}

An obvious desideratum for a joint state $\rho_{AB}$ is that it returns the correct marginal states for $A$ and $B$, i.e., $\mathrm{Tr}_B \rho_{AB} = \rho_A$ and $\mathrm{Tr}_A \rho_{AB} = \rho_B$. Given that $\rho_{AB} = \cs{A}{B} \star \rho_A$ by definition, and that 
$\rho_B = \mathrm{Tr}_A (\rho_A \cs{A}{B})$, as noted earlier, it follows that we require $\mathrm{Tr}_A (\cs{A}{B} \star \rho_A) = \mathrm{Tr}_A (\rho_A \cs{A}{B})$, hence also
\[
\mathrm{Tr}_{AB} (\cs{A}{B} \star \rho_A) = \mathrm{Tr}_{AB} (\rho_A \cs{A}{B}).
\]
This motivates the fourth axiom:
\[
  \textbf{Product when traced:} \ \ \tr (x \star y) = \tr(xy).
\]
All constructions considered in this work satisfy this requirement.

\subsection{Compositionality}

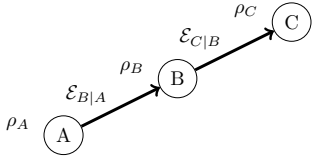
\begin{figure}[t]
    \scalebox{0.75}{\hspace{2cm}
    \begin{minipage}[c]{1.0\linewidth}
  \centering
  \[\begin{tikzpicture}
    \node[circle,draw] (a) at (0,0) {A};
    \node[circle,draw] (b) at (2,1) {B};
    \node[circle,draw] (c) at (4,2) {C};
    \draw[line width=1.5pt,->] (a) -- (b);
    \draw[line width=1.5pt,->] (b) -- (c);
    \draw (-.8,.2) node {$\rho_A$};
     \draw (.4,.7) node {$\mathcal{E}_{B|A}$};
    \draw (1.2,1.2) node {$\rho_B$};
    \draw (2.4,1.7) node {$\mathcal{E}_{C|B}$};
    \draw (3.2,2.2) node {$\rho_C$};
  \end{tikzpicture}\]
\end{minipage}
    }
\caption{Three  localized regions, $A$, $B$, and $C$, the state on $A$, $\rho_A$, the channel from $A$ to $B$, $\chan{A}{B}$, and the channel from $B$ to $C$, $\chan{B}{C}$.} \label{second}
\end{figure}

The final axiom for the star product concerns how it composes when we wish to describe a system at more than two distinct times. Consider the situation of Fig.~\ref{second}, with three  localized regions, labelled $A$, $B$ and $C$, joined by two channels, $\chan{A}{B}$ and $\chan{B}{C}$. 

There are two ways in which one can imagine forming the joint state $\rho_{ABC}$ on the composite of the three regions.  The first way proceeds as follows: form the joint state over the first two regions, $\rho_{AB}$; define a novel kind of channel, denoted $\chancomma{AB}{C}$, whose input is the composite of the first two regions and whose output is the third region; finally, take the star product of $\rho_{AB}$ with $\cscomma{AB}{C}$, the operator corresponding to $\chancomma{AB}{C}$: 
\be
\rho_{ABC} \equiv \cscomma{AB}{C} \star \rho_{AB}.
\label{FirstWay1}
\ee
The second way proceeds as follows: define a novel kind of channel, denoted $\chancomma{A}{BC}$,  that has the first region as input and the composite of the second and the third regions as output; next, take the star product of $\rho_{A}$ with $\cscomma{A}{BC}$, the operator corresponding to this channel: 
\be
\rho_{ABC} \equiv \cscomma{A}{BC} \star \rho_A.
\label{SecondWay1}
\ee

  We now consider the consequences, for both approaches,   of the fact that the only dependence that $C$ has on $A$ is mediated by $B$. 

In the first approach, this implies that the channel $\chancomma{AB}{C}$ is simply $\chan{B}{C}$, so that $\cscomma{AB}{C} = \cs{B}{C} \otimes \openone_A$.
Using this, together with the fact that $\rho_{AB} = \cs{A}{B} \star \rho_A$, Eq.~\eqref{FirstWay1} becomes
\be
\rho_{ABC} \equiv \cs{B}{C} \star (\cs{A}{B} \star \rho_{A}),
\label{FirstWay2}
\ee
where we have again adopted our convention of suppressing identity operators. In the second approach, this fact makes it natural to define the channel state for $\mathcal{E}_{BC|A}$ via the star product, as $\cscomma{A}{BC} \equiv \cs{B}{C} \star \cs{A}{B}$. This can be understood as a quantum analogue of the Markov condition for classical causal networks~\cite{pearl}. 

We have here allowed a channel state to appear as the second argument of the star product.  This is, however, no more objectionable than allowing a state over time to appear as the second argument, as we did in Eq.~\eqref{FirstWay1}. This is because, for every channel state $\cs{A}{B}$, there is a state over time $\rho_{AB}$ such that $\cs{A}{B} = d_A \rho_{AB}$ (where $d_A$ is the dimension of the Hilbert space of $A$).  To see that this is correct, we note that for channel state $\cs{A}{B}$ and input state $\frac{1}{d_A} \openone_A$, the state over time is $\rho_{AB} =  \cs{A}{B} \star \frac{1}{d_A} \openone_A$. Therefore, by our assumption of being product on commuting pairs, this reduces to $\rho_{AB} = \frac{1}{d_A}  \cs{A}{B}$. Invoking convex-bilinearity, we can therefore write $\cs{B}{C} \star \cs{A}{B}$ as $d_A (\cs{B}{C} \star \rho_{AB})$. 

Using $\cscomma{A}{BC} \equiv \cs{B}{C} \star \cs{A}{B}$, Eq.~\eqref{SecondWay1} then becomes
\be
\rho_{ABC} \equiv (\cs{B}{C} \star \cs{A}{B}) \star \rho_A.
\label{SecondWay2}
\ee

Our last condition on the star product is motivated by the idea that 
the two techniques just described for defining a joint state over three times should give the same result, 
\be
 \cs{B}{C} \star (\cs{A}{B} \star \rho_{A})  = (\cs{B}{C} \star \cs{A}{B}) \star \rho_{A}.
\ee
This can be ensured by taking the star product to be associative (note that the considerations above merely motivate associativity but do not imply it). 
Our final condition, therefore, is
\[
  \mathbf{Associativity:} \ \ x \star (y \star z) = (x \star y) \star z.
\]

The W formulation defines an associative star product. The LS star product, however, is not associative, as noted by LS in Sec.~VII of \cite{leiferspekkens:bayesian}. In particular, the second argument of the LS star product is required to be a positive operator---in order for the square root of the operator to be uniquely defined---and yet nonpositive operators arise as joint states over time. Indeed, the fact that the LS star product does not support composition across time was highlighted by LS as one of its most significant shortcomings, motivating the search for alternative approaches. 

The FJV construction is interesting with regard to associativity. One can verify that the FJV construction for a single qubit at three times gives a joint state:
\[
\rho^{(\mathrm{FJV})}_{ABC} = \cs{B}{C} \star_{\mathrm{FJV}} (\cs{A}{B} \star_{\mathrm{FJV}} \rho_A),
\]
where, as noted previously, $\star_{\mathrm{FJV}}$ is the Jordan product.
Although the Jordan product is non-associative in general,   it is associative for triples of operators of the form $\cs{B}{C} \otimes \openone_A, \cs{A}{B}\otimes \openone_B$, and $\rho_A\otimes \openone_B \otimes \openone_C$, that is, (suppressing identities as usual) 
\begin{equation}\label{assjord}
\cs{B}{C} \star_{\mathrm{FJV}} (\cs{A}{B} \star_{\mathrm{FJV}} \rho_A) = (\cs{B}{C} \star_{\mathrm{FJV}} \cs{A}{B}) \star_{\mathrm{FJV}} \rho_A.
\end{equation}
The reason is that $\rho_A\otimes \openone_B \otimes \openone_C$ commutes with $\cs{B}{C} \otimes \openone_A$. These observations extend to a single qubit considered over an arbitrary number of time steps. Hence although the Jordan product is non-associative, it \emph{is} associative when restricted to operators of the correct form to describe a quantum system passing through a sequence of channels.  Hence a unique joint state is defined without privileging any particular way of grouping the different time steps.   When one considers an arbitrary number of qubits, the situation becomes more subtle: an FJV star product can again be inferred, but in this case it is not the Jordan product of Eq.~(\ref{fjvstar}).  A similar situation arises for systems whose dimension is a power of a prime.  For other systems, the proposal is ambiguous.  We omit the details as these issues are non-essential for our purposes.

\section{Main theorem}\label{sec:main}

We now have five criteria for states across time, expressed as axioms for a star product: Hermiticity, convex-bilinearity, product when commuting, product when traced, and associativity. The satisfaction or otherwise of these axioms, for each of LS, FJV, and the W 
construction are summarised in Table~\ref{table}. 
\begin{table*}[t]
  \begin{center}
    \begin{tabular}{| l || c | c | c | c | c |}
    \hline
     &  & Convex- & Product & Product  &   \\
          & Hermitian? & bilinear? & on commuting pairs? & when traced? &  Associative? \\
     \hline \hline
    LS & \checkmark & $\times$ & \checkmark & \checkmark  & $\times$ \\\hline
    FJV & \checkmark & \checkmark & \checkmark & \checkmark &  $\times^\star$\\\hline
    W & \checkmark & \checkmark & $\times$ & \checkmark & \checkmark\\
    \hline
    \end{tabular}
  \end{center}
  \caption{Satisfaction of axioms by star products of the three constructions of states over time. ($^\star$see text for details)}\label{table}
\end{table*}
None of these constructions satisfies all of the axioms. 
\begin{theorem}\label{maintheorem}
Let $\H_n$ be the set of $n$-by-$n$ Hermitian matrices. There is no function $\star \colon \H_n \times \H_n \to \H_n$ satisfying {\em convex-bilinearity}, {\em product on commuting pairs}, {\em product when traced}, and {\em associativity}.
\end{theorem}
In other words, there is no definition of a state over time that satisfies the five axioms given in this paper. We provide the proof of this result below.

It is natural to consider what happens when one or more of the axioms is dropped. 

Suppose that Hermiticity is not assumed, and that the star product is assumed to be a map $\star \colon \M_n \times \M_n \to \M_n$, where $\M_n$ is the complex vector space of all $n$-by-$n$ matrices. In this case, as we also show below, the remaining four axioms force the following form of the star product.
\begin{theorem}\label{nonhermitiantheorem}
Let $\M_n$ be the complex vector space of $n$-by-$n$ matrices. The only functions $\star \colon \M_n \times \M_n \to \M_n$ satisfying convex-bilinearity, product on commuting pairs, product when traced, and associativity are $x \star y=xy$ and $x \star y = yx$.
\end{theorem}
In this case, there are only two possibilities for the star product, corresponding to straightforward (left and right) matrix multiplication.   The possibility of taking the star product to be matrix multiplication was noted in \cite{leiferspekkens:bayesian} as one way to overcome the shortcomings of the LS star product. We comment on some of the consequences of such a choice in the discussion section.   

If associativity is dropped, on the other hand, with the other four axioms (including Hermiticity) maintained, then the FJV construction provides an interesting example of a means to construct states over time.

Finally, in Appendix~\ref{sec:largerspaces}, we also consider the possibility that the result of the star product is a matrix in a larger (or smaller) state space, with 
\[
\star: \M_n \times \M_n \to \M_m \qquad m\neq n.
\]
With suitable refinements for the axioms of {\em product on commuting pairs} and {\em associativity}, the conclusion of the theorem remains essentially the same.

\subsection{Derivation of main theorems}

Key to the derivation of Theorems~\ref{maintheorem} and \ref{nonhermitiantheorem} is the following result \cite[Theorem~1]{heunenhorsman:matrixmultiplication}. Using conditions on a function $\star \colon \M_n \times \M_n \to \M_n$, given as
\begin{enumerate}
  \item[(A)] ``\emph{Associativity}'': $x \star (y \star z) = (x \star y) \star z$ for all $x,y,z \in \A$;
  \item[(B)] ``\emph{Bilinearity}'': $(\lambda x) \star y = \lambda(x \star y) = x \star (\lambda y)$, $(x+y) \star z = (x \star z) + (y \star z)$, and $x \star (y+z) = (x \star y) + (x \star z)$ for scalars $\lambda$ and $x,y,z \in \A$;
  \item[(I)] ``\emph{Identity}'': $x \star 1 = x$ for all $x \in \A$, where $1 \in \A$ is the identity matrix.
  \item[(T)] ``\emph{Trace}'': $\tr(x \star y) = \tr(xy)$ for all $x,y \in \A$.
  \item[(O)] ``\emph{Orthogonality}'': $x \star y = 0$ when $xy=yx=0$, $xx=x$, and $yy=y$, \\for $x,y \in \A$ that have rank one.
\end{enumerate}

\noindent the result is:

\begin{quote}
  The function $\star \colon \M_n \times \M_n \to \M_n$ satisfies (A), (B), (I), (T) and (O) if and only if either $x \star y=xy$ for all $x,y \in \M_n$, or $x \star y = yx$ for all $x,y \in \M_n$.    
\end{quote}
Here, condition (A) is precisely our associativity condition and (T) is the product when traced axiom. Condition (I) is the identity composition, Eq.~\eqref{identity} above, and condition~(O) is that
\be
  x \star y = 0 \text{ when } xy=yx=0,\ xx=x, \text{ and }yy=y, 
  \text{ for } x,y \in \M_n.
\ee
Both the condition (I) and the condition (O) follow straightforwardly from the product on commuting pairs axiom. The one remaining condition, (B), is bilinearity over the complex numbers. On the face of it this is slightly stronger than our axiom of convex-bilinearity. Nonetheless, Theorems~\ref{maintheorem} and \ref{nonhermitiantheorem} follow from \cite[Theorem~1]{heunenhorsman:matrixmultiplication}, along with the following lemma.
\begin{lemma}\label{reductionlemma}
Let $\H_n$ be the set of $n$-by-$n$ Hermitian matrices, and $\M_n$ the set of all $n$-by-$n$ complex matrices. Any convex-bilinear function $\star: \H_n \times \H_n \to \H_n$ satisfying product on commuting pairs, product when traced, and associativity, extends uniquely to a complex bilinear function $\star: \M_n \times \M_n \to \M_n$, which also satisfies product on commuting pairs, product when traced, and associativity.
\end{lemma}
Theorem~ \ref{nonhermitiantheorem} follows from \cite[Theorem~1]{heunenhorsman:matrixmultiplication}, since if the star product is merely convex-bilinear and satisfies the other requirements, then it follows from the lemma that it must in fact be complex bilinear and still satisfy the other requirements. Theorem~\ref{maintheorem} follows since if there were a function on $\H_n \times \H_n$ satisfying the five axioms, then by the lemma it would extend to a function on $\M_n \times \M_n$ that satisfies (A), (B), (I), (T) and (O), hence could only be left or right matrix multiplication: but matrix multiplication does not preserve hermiticity.

We now provide the proof of lemma ~\ref{reductionlemma}.
\begin{proof}
  We first show that any convex-bilinear function $\star: \H_n \times \H_n \to \H_n$ satisfying the other axioms is in fact bilinear over the reals. It follows from the product on commuting pairs axiom that $x \star 0 = 0 \star x = 0$   where $0$ here denotes the $n$-by-$n$ matrix all of whose components are 0 . By considering convex combinations of the form $x = r x'+(1-r) 0$ (for $0\leq r\leq 1$), or $x = (1/r) x' + (1-1/r) 0$ (for $r\geq 1$), it is then easy to show that $(r x ) \star y = r (x \star y)$ for any $r \geq 0$. Recall that a \emph{conic combination} of $x_i \in \H_n$ is $\sum_i r_i x_i$ for nonnegative real weights $r_i$. Writing $r = \sum_i r_i$ yields $(\sum_i r_i x_i) \star y = r ((1/r \sum_i r_i x_i ) \star y) = \sum_i r_i (x_i \star y)$, where the last equality follows from convex bilinearity. Hence the function $\star$ is linear over conic combinations. Finally, consider a linear combination $x=\sum_{i \in I} s_i x_i$ with $s_i \in \mathbb{R}$. Chop the index set $I$ into $I_- = \{ i \in I \mid s_i < 0\}$ and $I_+ = \{ i \in I \mid s_i \geq 0 \}$. Then $x + \sum_{i \in I_-} |s_i|x_i = \sum_{i \in I_+} s_i x_i$, where each side of this equation is a conic combination of elements of $\H_n$. Taking the star product with $y$ yields $(x + \sum_{i \in I_-} |s_i|x_i ) \star y = (\sum_{i \in I_+} s_i x_i) \star y$, which after using conic linearity and rearranging gives $x\star y = \sum_i s_i (x_i \star y)$.
Similar reasoning applied to the second argument yields bilinearity over the reals, as claimed.

We now prove that any product $\star \colon \H_n \times \H_n \to \H_n$, which is bilinear over the reals, and satisfies the remaining axioms, extends to a product on $\M_n$, which is bilinear over the complex numbers, and still satisfies the remaining axioms. Recall that a matrix $x \in \M_n$ is Hermitian if $x^\dag = x$, and anti-Hermitian if $x^\dag = -x$; there is a bijection $x \mapsto ix$ between Hermitian and anti-Hermitian matrices. Any matrix $x \in \M_n$ equals $x=x_h+x_a$ for a unique Hermitian $x_h \in \H_n$ and anti-Hermitian $x_a$, given by $x_h=\tfrac{1}{2}(x+x^\dag)$ and $x_a=\tfrac{1}{2}(x-x^\dag)$. Define $\star \colon \M_n \times \M_n \to \M_n$ by 
\[
  x \star y = x_h \star y_h - i(x_h \star iy_a) -i(ix_a \star y_h) -(ix_a \star iy_a).
\]
Observe that this is well-defined, and that
\begin{align*}
  (x \star y)_h & = x_h \star y_h - ix_a \star iy_a,  \\
  (x \star y)_a & = -i(x_h \star iy_a + ix_a \star y_h). 
\end{align*} 
Associativity of the extended function $\M_n \times \M_n \to \M_n$ now follows from associativity of the original function $\H_n \times \H_n \to \H_n$ by a straightforward computation, and the same goes for the condition of being product when traced and for complex bilinearity. We write out the latter explicitly. Suppose that $\lambda = \alpha + i\beta$ for $\alpha,\beta \in \R$. Observe that $(\lambda x)_h = \alpha x_h + \beta i x_a$, and $i(\lambda x)_a = \alpha i x_a - \beta x_h$. Then:
\begin{align*}
    (\lambda x) \star y
    & = (\alpha x_h + \beta ix_a) \star y_h - i((\alpha x_h + \beta i x_a) \star iy_a) \\
    & \phantom{ = }-i((\alpha ix_a-\beta x_h) \star y_h) - (\alpha i x_a - \beta x_h) \star iy_a \\
    & = (\alpha +\beta i) (x_h \star y_h) -i (\alpha+\beta i) (x_h \star (iy_a)) \\
    & \phantom{ = } -i (\alpha+\beta i) ((ix_a) \star y_h) - (\alpha+\beta i) (ix_a \star iy_a) \\
    & = \lambda (x \star y),
\end{align*}
and similarly $x \star (\lambda y) = \lambda (x \star y)$.
Using $(x+y)_h = x_h + y_h$ and $(x+y)_a = x_a + y_a$ gives $(x+y) \star z = x \star z + y \star z$ similarly easily.

Finally, consider the condition of being product on commuting pairs.

First, consider the case where $x \in \M_n$ is \emph{normal}, that is, $xx^\dag = x^\dag x$. 
By assumption $xy=yx$, and so $x^\dag y^\dag = (yx)^\dag = (xy)^\dag = y^\dag x^\dag$, too.
By Fuglede's theorem~\cite{fuglede:commutativity}, also $xy^\dag = y^\dag x$, and so $x^\dag y = y x^\dag$. Therefore
\begin{align*}
    x_hy_h 
    & = 1/4 (xy+x^\dag y + xy^\dag + x^\dag y^\dag ) \\
    & = 1/4 (yx+y x^\dag + y^\dag x + y^\dag x^\dag )\\
    & = y_h x_h\text{,}
\end{align*}
and similarly $x_hy_a=y_ax_h$, $x_ay_h=y_hx_a$, and $x_ay_a = y_ax_a$.
It now follows that $x_h \star y_h = x_hy_h$, $x_h \star iy_a = ix_hy_a$, $ix_a \star y_h = ix_ay_h$, and $ix_a \star iy_a = -x_ay_a$. 
But   this, together with the bilinearity of $\star : \M_n \times \M_n \to \M_n$, implies: 
\begin{align*}
    x \star y
    & = x_h \star y_h - i(x_h \star iy_a + ix_a \star y_h) - (ix_a \star iy_a)\\
    & = x_h y_h + x_h y_a + x_a y_h + x_a y_a \\
    & = xy.
\end{align*}

  Now we consider the case of a general $x \in \M_n$ (not necessarily normal).  
  Because both hermitian and anti-hermitian matrices are normal, an arbitrary matrix $x \in \M_n$ can be written as a sum of two normal matrices $x=x_h+x_a$. Combining   the result for normal matrices   with bilinearity of the extended function $\star : \M_n \times \M_n \to \M_n$, we infer that $x \star y = xy$ for any $y \in \M_n$.

  This concludes the proof of Lemma~\ref{reductionlemma}.
\end{proof}

\section{Conclusions}\label{sec:discussion}

We have considered methods for constructing quantum states over time, under the assumption that the state over time is an operator on the tensor product of Hilbert spaces associated with the temporally localised subsystems. The motivation was to explore the possibilities for a quantum treatment of time-like separated systems that is as close as possible to the standard quantum treatment of composite systems at a single time. In particular, given the well developed analogy between density matrices and probability distributions -- and the fact that classical joint probabilities can be defined regardless of the spatio-temporal relationships between a set of events -- it is interesting to see whether this analogy can be extended, or whether it is fundamentally limited in scope by an asymmetric treatment of space and time by quantum theory. We focused on simple cases of spatio-temporally localized quantum systems evolving through a sequence of quantum channels. We identified a set of physically-motivated criteria for such a construction, and Theorem~\ref{maintheorem} shows that there is no construction that satisfies all of the criteria. 

One possible conclusion that may be drawn from Theorem~\ref{maintheorem}, then, is that there is something fundamentally misguided in attempting to treat quantum systems over time in the same manner as composite systems at a single time. Interpreted thus, Theorem~\ref{maintheorem} might be seen, for example, as lending support to the view that a quantum system localized in time should be described with two Hilbert spaces, representing an incoming and outgoing state, as per the multi-time \cite{aharonov2007two, aharonov2009multiple, aharonov2013each, silva2014pre}, quantum combs \cite{ChiribellaEtAl_2009, Chiribella_2012, ChiribellaEtAl_2013} and process matrices \cite{Oreshkov2012, AraujoEtAl_2014, Costa2016} formalisms. 

But one should also ask if there is a way that the apparent no-go result of Theorem~\ref{maintheorem} can be avoided. The most obvious way is to question whether the axioms that we have identified are themselves reasonable. The axioms of {\em convex-bilinearity} and of {\em product when traced} seem to be essential if the joint state is to play a role that is analogous to the role played by the joint state of composite systems at a given time. Reproducing classical joint probabilities when the initial state and the channel states all commute also seems essential, at least if the quantum states over time are to be seen as generalizations of classical probability distributions in the way that standard quantum density matrices are seen as generalizations of classical probability distributions. However, it is reasonable to judge a pairing of initial state and channel state to be classical only when they are jointly diagonalisable {\em in a basis that is a tensor product of a fixed basis for each system} rather than simply jointly diagonalisable.  Therefore, the assumption that the joint state should be equivalent to a joint probability distribution for such pairs only warrants that the star product should reduce to the matrix product on certain pairs of commuting operators, rather than all pairs of commuting operators, as the condition asserts.    Hence there may be some wiggle room in constructing joint states that reproduce classical joint probabilities in the desired sense but do not satisfy the axiom. 

As noted in the previous section, if the state over time is not assumed to be Hermitian, then Theorem 2 shows that matrix multiplication is the only construction that satisfies the remaining axioms. We also noted that if one does not assume associativity, then the FJV construction satisfies the other four axioms.  Furthermore, although the FJV star product is not associative for arbitrary triples of operators, we saw that on the restricted domain of triples of operators corresponding to a sequence of channel states, it {\em is} associative.  
A disadvantage of the FJV construction as stated, however, is that it is defined only for collections of qubits. It can be applied to systems   whose dimension is a power of 2 by regarding them as composite systems made up of qubits, but then there is an arbitrariness in the choice of how to factorize the larger Hilbert space into component systems.   There is also a natural generalization of the proposal from a qubit to a system of prime dimension, but how to generalize the proposal to systems of arbitrary dimension remains ambiguous. Note that there is an interesting relationship between the FJV construction and that corresponding to matrix multiplication, at least for a single qubit considered at two times: the joint state $\rho^{(\mathrm{FJV})}_{AB}$ is a Hermitian operator whose eigenspectrum coincides with the positive and real eigenspectrum of $\rho^{(\mathrm{MM})}_{AB} = \cs{A}{B} \rho_A$. That is, $\rho^{(\mathrm{FJV})}_{AB}$ can be obtained from $\rho^{(\mathrm{MM})}_{AB} $ by removing the latter's imaginary eigenvalues.

In all of the proposals for defining a state over time that we have considered, the set of operators that can describe a joint state of a system considered at two times is {\em distinct} from the set of operators that can describe a state of two copies of the system at a given time.  This contrasts with classical probability theory, where the possible joint probability distributions on a pair of systems is the same in the two cases.  This suggests that even if a sensible notion of a state over time can be defined, quantum theory still manifests an asymmetry between space and time that is not present in classical probability theory (analogous, perhaps, to the fact that relativity theory still distinguishes space and time through the signature of the metric).  Indeed, this conclusion is supported by the results of \cite{Ried2015}, which demonstrate that in quantum theory, unlike classical theories, the set of correlations that can be generated by two systems that are related as cause and effect is different from the set that can be generated by two systems that are both effects of a common cause.
Further investigations into operationally-motivated definitions of states over time are likely to shed additional light on the precise nature of this asymmetry.   

Finally, in seeking to define a state for a composite over time, we have here considered to what extent the formal properties of such a state could resemble those of a state for a composite at a given time. It would also be interesting to approach the problem from a purely operational point of view, where the aim is to define a joint state by reference to predictions for experimental interventions on systems at different times. For instance, it would be interesting to consider the relationships between the various constructions considered here and the predictions for the outcomes of weak measurements performed across multiple time steps \cite{marcovitchreznik}.


\def\contribs#1{{\vskip5.5pt\noindent {\fontsize{9}{11}\selectfont Acknowledgements.}\fontsize{8}{11}\selectfont\enskip #1}}
\contribs{This research was supported in part by Perimeter Institute for Theoretical Physics. Research at Perimeter Institute is supported by the Government of Canada through the Department of Innovation, Science and Economic Development Canada and by the Province of Ontario through the Ministry of Research, Innovation and Science. DH and JB acknowledge support from an FQXi Large Grant, and from the NQIT Quantum Hub. DH was also supported by EPSRC grant EP/L022303/1. This project/publication was made possible in part through the support of a grant  from the John Templeton Foundation. The opinions expressed in this publication are those of the author(s) and do not necessarily reflect the views of the John Templeton Foundation.  CH was supported by EPSRC Fellowship EP/L002388/1.
}

\appendix
\section{The FJV construction as a star product}\label{sec:fjv}

In this section, we show that for a qubit at two times, the Fitzsimons-Jones-Vedral construction \eqref{fjv} corresponds to the Jordan product, $\rho_{AB} = \frac12\left( \rho_A \cs{A}{B} + \cs{A}{B}\rho_A \right)$.
Recall that the FJV construction defines a state of two systems as
\be \rho^{(\mathrm{FJV})}_{AB} = \frac{1}{4} \left( \sum_{i=0}^{3}\av{\sigma_i \otimes \sigma_j} \ \sigma_i \otimes \sigma_j  \right). \label{fjv_repeat}\ee
The expectations are defined operationally with respect to measurements of the Pauli operators with the projection postulate as the state update rule.  Thus, $\av{\sigma_i \otimes \sigma_j}$ is read for a state over time as 
the expectation value of the product of outcomes of a projective measurement of $\sigma_i$  followed by a measurement of $\sigma_j$ (where, after the first measurement, $A$ is in the eigenstate of $\sigma_i$ corresponding to the outcome of that measurement).

Using the fact that the state at the output of the channel is given by $\tr_A(\cs{A}{B}\rho_A)$, we have
\begin{align} 
  \av{\sigma_i \otimes \sigma_j} 
  & = \tr\left(\cs{A}{B}(\rho_A^{i+} \otimes \sigma_j)\right) - \tr\left(\cs{A}{B}(\rho_A^{i-} \otimes \sigma_j)\right) \notag\\
  & = \tr\left(\cs{A}{B}( (\rho_A^{i+} - \rho_A^{i-}) \otimes \sigma_j)\right), 
\end{align}
where $\rho_A^{i+}$ is the subnormalized state of $A$ after projection onto the $+1$ eigenspace of $\sigma_i$, and similarly for $\rho_A^{i-}$, i.e., $\rho_A^{i\pm} \equiv P^{i\pm}\rho_A P^{i\pm}$ where $P^{i\pm} \equiv (\openone \pm \sigma_i)/2$. 

Noting that
\begin{align} 
  & \rho_A^{i+} - \rho_A^{i-} \notag \\
  & = \frac14\left((\openone + \sigma_i)\rho_A (\openone + \sigma_i) - (\openone - \sigma_i)\rho_A(\openone - \sigma_i)\right) \notag \\
  & = \frac12(\sigma_i\rho_A + \rho_A\sigma_i), 
\end{align}
we obtain
\begin{align} 
 & \av{\sigma_i \otimes \sigma_j} \notag \\
 & = \frac12 \tr\left( \cs{A}{B}((\sigma_i\rho_A + \rho_A\sigma_i) \otimes \sigma_j) \right) \notag \\ 
 & = \frac12 \tr\left(\rho_A \cs{A}{B}(\sigma_i \otimes \sigma_j) + \cs{A}{B}\rho_A (\sigma_i \otimes \sigma_j)\right), \label{exp1}
\end{align}
where in the last equality we have used the cyclic property of the trace.
 
Eq.~\eqref{fjv_repeat} implies that 
\be \label{exp2} \av{\sigma_i \otimes \sigma_j} = \tr\left(\rho_{AB}^{(\mathrm{FJV})} (\sigma_i \otimes \sigma_j)\right) , \ee
Combining Eqs.~\eqref{exp1} and~\eqref{exp2}, and using the fact that the expectation values of products of Paulis uniquely define an operator, 
we conclude that $\rho_{AB}^{(\mathrm{FJV})} = \frac12(\rho_A \cs{A}{B} + \cs{A}{B}\rho_A)$ as claimed.

\section{Larger state spaces}\label{sec:largerspaces}

One could consider functions $\star \colon \M_n \times \M_n \to \M_m$ for $m$ possibly different from $n$. The condition of convex-bilinearity then still makes sense, but associativity and being product on commuting pairs do not. The most natural solution is to require that there exists a linear function $p \colon \M_m \to \M_n$ and adapt the criteria accordingly:
\begin{itemize}
  \item \textbf{associativity}: $x \star p(y \star z) = p(x \star y) \star z$;
  \item \textbf{product on commuting pairs}: if $xy=yx$ then $p(x \star y)=xy$;
  \item \textbf{product when traced}: $\tr(p(x \star y)) = \tr(xy)$.
\end{itemize}
But then one easily sees that $p \circ \star \colon \M_n \times \M_n \to \M_n$ satisfies the original criteria for a star product. So by Theorem~\ref{nonhermitiantheorem}, either $p(x \star y)=xy$ for all $x,y \in \M_n$ or $p(x \star y) = yx$. Hence $p(x \star \openone)=x$ and it follows that $\mathrm{rank}(p) \geq n^2$ and so $m \geq n$. Thus the function $j \colon \M_n \to \M_m$ given by $j(x)=x \star \openone$ satisfies $p(j(x))=x$.
But then $\tr(x \star y) = \tr(j(xy))$, so that $x \star y$ and $j(xy)$ result in the same probabilistic predictions when measuring a state. In other words, there is nothing to be gained by moving from $\M_n$ to $\M_m$ for $m>n$.

\bibliographystyle{plain}
\bibliography{starproducts}

\begin{thebibliography}{10}

\bibitem{aharonov2013each}
Yakir Aharonov, Sandu Popescu, and Jeff Tollaksen.
\newblock Each instant of time a new universe.
\newblock {\em Quantum Theory: A Two-Time Success Story: Yakir Aharonov
  Festschrift}, page~21, 2013.

\bibitem{aharonov2009multiple}
Yakir Aharonov, Sandu Popescu, Jeff Tollaksen, and Lev Vaidman.
\newblock Multiple-time states and multiple-time measurements in quantum
  mechanics.
\newblock {\em Physical Review A}, 79(5):052110, 2009.

\bibitem{aharonov2007two}
Yakir Aharonov and Lev Vaidman.
\newblock The two-state vector formalism: An updated review.
\newblock {\em Time in Quantum Mechanics}, pages 399--447, 2007.

\bibitem{AraujoEtAl_2014}
Mateus Ara\'ujo, Fabio Costa, and \v{C}aslav Brukner.
\newblock Computational advantage from quantum-controlled ordering of gates.
\newblock {\em Phys. Rev. Lett.}, 113:250402, Dec 2014.

\bibitem{Chiribella_2012}
Giulio Chiribella.
\newblock Perfect discrimination of no-signalling channels via quantum
  superposition of causal structures.
\newblock {\em Physical Review A}, 86(4):040301, 2012.

\bibitem{ChiribellaEtAl_2009}
Giulio Chiribella, Giacomo~Mauro D'Ariano, and Paolo Perinotti.
\newblock Theoretical framework for quantum networks.
\newblock {\em Phys. Rev. A}, 80(2):022339, 2009.

\bibitem{ChiribellaEtAl_2013}
Giulio Chiribella, Giacomo~Mauro D'Ariano, Paolo Perinotti, and Benoit Valiron.
\newblock Quantum computations without definite causal structure.
\newblock {\em Physical Review A}, 88(2):022318, 2013.

\bibitem{choi}
Man-Duen Choi.
\newblock Completely positive linear maps on complex matrices.
\newblock {\em Linear Algebra and its Applications}, 10:285--290, 1975.

\bibitem{Costa2016}
Fabio Costa and Sally Shrapnel.
\newblock {Quantum causal modelling}.
\newblock {\em New Journal of Physics}, 18(6):063032, June 2016.

\bibitem{feynman}
Richard~P. Feynman.
\newblock Space-time approach to non-relativistic quantum mechanics.
\newblock {\em Reviews of Modern Physics}, 20(2):367, 1948.

\bibitem{fitzsimonsjonesvedral:causation}
Joseph Fizsimons, Jonathan~A. Jones, and Vladko Vedral.
\newblock Quantum correlations which imply causation.
\newblock {\em arXiv:1302.2731}, 2013.

\bibitem{fuglede:commutativity}
Bent Fuglede.
\newblock A commutativity theorem for normal operators.
\newblock {\em Proceedings of the National Academy of Sciences of the USA},
  36(1):35--40, 1950.

\bibitem{GellmannHartle1990}
Murray Gell-Mann and James~B Hartle.
\newblock Quantum mechanics in the light of quantum cosmology.
\newblock {\em Complexity, entropy and the physics of information}, 8, 1990.

\bibitem{griffiths1984consistent}
Robert~B Griffiths.
\newblock Consistent histories and the interpretation of quantum mechanics.
\newblock {\em Journal of Statistical Physics}, 36(1-2):219--272, 1984.

\bibitem{Gross}
David Gross.
\newblock Hudson's theorem for finite-dimensional quantum systems.
\newblock {\em Journal of Mathematical Physics}, 47(12):122107, 2006.

\bibitem{gutoskiwatrous:games}
Gus Gutoski and John Watrous.
\newblock Toward a general theory of quantum games.
\newblock In {\em STOC}, pages 565--574, 2007.

\bibitem{hardy2007towards}
Lucien Hardy.
\newblock Towards quantum gravity: a framework for probabilistic theories with
  non-fixed causal structure.
\newblock {\em Journal of Physics A: Mathematical and Theoretical},
  40(12):3081, 2007.

\bibitem{Hardy3385}
Lucien Hardy.
\newblock The operator tensor formulation of quantum theory.
\newblock {\em Philosophical Transactions of the Royal Society of London A:
  Mathematical, Physical and Engineering Sciences}, 370:3385--3417, 2012.

\bibitem{heunenhorsman:matrixmultiplication}
Chris Heunen and Clare Horsman.
\newblock Matrix multiplication is determined by orthogonality and trace.
\newblock {\em Linear Algebra and its Applications}, 439(12):4130--4134, 2013.

\bibitem{Jamiolkowski1972}
A.~Jamio{\l}kowski.
\newblock Linear transformations which preserve trace and positive
  semidefiniteness of operators.
\newblock {\em Rep. Math. Phys.}, 3:275--278, 1972.

\bibitem{leifer2006quantum}
Matthew~S Leifer.
\newblock Quantum dynamics as an analog of conditional probability.
\newblock {\em Physical Review A}, 74(4):042310, 2006.

\bibitem{leifer}
Matthew~S. Leifer and David Poulin.
\newblock Quantum graphical models and belief propagation.
\newblock {\em Annals of Physics}, 323(8):1899--1946, 2008.

\bibitem{leiferspekkens:bayesian}
Matthew~S. Leifer and Robert~W. Spekkens.
\newblock Towards a formulation of quantum theory as a causality neutral theory
  of {B}ayesian inference.
\newblock {\em Physical Review A}, 88(5):052130, 2013.

\bibitem{marcovitchreznik}
Shmuel Marcovitch and Benni Reznik.
\newblock Structural unification of space and time correlations in quantum
  theory.
\newblock {\em arXiv:1103.2557}, 2011.

\bibitem{omnes1988logical}
Roland Omnes.
\newblock Logical reformulation of quantum mechanics. {I}. {F}oundations.
\newblock {\em Journal of Statistical Physics}, 53(3-4):893--932, 1988.

\bibitem{oreshkovcerf:withouttime}
Ognyan Oreshkov and Nicolas~J. Cerf.
\newblock Operational quantum theory without predefined time.
\newblock {\em New Journal of Physics}, 18:073037, 2016.

\bibitem{Oreshkov2012}
Ognyan Oreshkov, Fabio Costa, and \v{C}aslav Brukner.
\newblock Quantum correlations with no causal order.
\newblock {\em Nature communications}, 3:1092, 2012.

\bibitem{pearl}
Judea Pearl.
\newblock {\em Causality: Models, Reasoning, and Inference}.
\newblock Cambridge University Press, New York, NY, USA, 2000.

\bibitem{Ried2015}
Katja Ried, Megan Agnew, Lydia Vermeyden, Dominik Janzing, Robert~W. Spekkens,
  and Kevin~J. Resch.
\newblock {A quantum advantage for inferring causal structure}.
\newblock {\em Nature Physics}, 11(5):414--420, March 2015.

\bibitem{silva2014pre}
Ralph Silva, Yelena Guryanova, Nicolas Brunner, Noah Linden, Anthony~J Short,
  and Sandu Popescu.
\newblock Pre-and postselected quantum states: Density matrices, tomography,
  and kraus operators.
\newblock {\em Physical Review A}, 89(1):012121, 2014.

\bibitem{Sorkin1997}
Rafael~D. Sorkin.
\newblock {Forks in the road, on the way to quantum gravity}.
\newblock {\em International Journal of Theoretical Physics},
  36(12):2759--2781, December 1997.

\bibitem{wooters:discretewigner}
William~K. Wooters.
\newblock A {W}igner-function formulation of finite-state quantum mechanics.
\newblock {\em Annals of Physics}, 176:1--21, 1987.

\end{thebibliography}

\end{document}